\documentclass[prx,aps,twocolumn,showpacs,superscriptaddress,nofootinbib]{revtex4-1}
\usepackage[colorlinks=true,citecolor=blue,linkcolor=magenta]{hyperref}
\usepackage{graphicx}
\usepackage{dcolumn}
\usepackage{bm}
\usepackage[bbgreekl]{mathbbol}
\usepackage{amsmath,dsfont}
\usepackage{color}
\usepackage{soul}
\usepackage{changes}

\begin{document}

\title{Packaged photonic chip-based soliton microcomb using an ultralow-noise laser}

\author{Arslan S. Raja}
\thanks{These authors contributed equally to this work.}
\affiliation{Swiss Federal Institute of Technology Lausanne (EPFL), CH-1015 Lausanne, Switzerland}

\author{Junqiu Liu}
\thanks{These authors contributed equally to this work.}
\affiliation{Swiss Federal Institute of Technology Lausanne (EPFL), CH-1015 Lausanne, Switzerland}

\author{Nicolas Volet}
\affiliation{University of California, Santa Barbara (UCSB), CA 93106, USA}

\author{Rui Ning Wang}
\affiliation{Swiss Federal Institute of Technology Lausanne (EPFL), CH-1015 Lausanne, Switzerland}

\author{Jijun He}
\affiliation{Swiss Federal Institute of Technology Lausanne (EPFL), CH-1015 Lausanne, Switzerland}
\author{Erwan Lucas}
\affiliation{Swiss Federal Institute of Technology Lausanne (EPFL), CH-1015 Lausanne, Switzerland}

\author{Romain Bouchand}
\affiliation{Swiss Federal Institute of Technology Lausanne (EPFL), CH-1015 Lausanne, Switzerland}

\author{Paul Morton}
\affiliation{Morton Photonics, West Friendship, MD 21794, USA}

\author{John Bowers}
\affiliation{University of California, Santa Barbara (UCSB), CA 93106, USA}
\author{Tobias J. Kippenberg}
\email[]{tobias.kippenberg@epfl.ch}
\affiliation{Swiss Federal Institute of Technology Lausanne (EPFL), CH-1015 Lausanne,
Switzerland}

\date{\today}
\begin{abstract}
Photonic chip-based soliton microcombs have shown rapid progress and have already been used in many system-level applications, including coherent communications, astrophysical spectrometer calibration and ultrafast ranging. While there has been substantial progress in realizing soliton microcombs that rely on compact laser diodes, culminating in devices that only utilize a semiconductor amplifier or a self-injection-locked laser as a pump source, accessing and generating single soliton states with electronically detectable line rates from a compact laser module has remained challenging. Here we demonstrate a current-initiated,  $\mathrm{Si_3N_4}$ chip-based, 99-GHz soliton microcomb driven directly by an ultra-compact, low-noise laser. This approach does not require any fast laser tuning mechanism, and single soliton states can be accessed by changing the current of the laser diode.  A simple, yet reliable, packaging technique has been developed to demonstrate the viability of such a  microcomb system in field-deployable applications. 

\end{abstract}
\maketitle
The development of optical frequency combs led to crucial scientific and technological advancements in the field of optical metrology and spectroscopy \cite{Udem:02}.  Microcombs allow optical frequency combs to be generated using integrated microresonators with broad bandwidth, ultra-low power and repetition rates in  GHz to THz range, and have developed into an active research area at the intersection of frequency metrology, integrated photonics and nonlinear soliton dynamics \cite{DelHaye:07, Kippenberg:18}.
Microcombs are compatible with wafer scale mass manufacturing and on-chip applications. 
 A particularly important phenomena in the field of Kerr combs are dissipative Kerr solitons (DKS), which constitute continuously circulated ultrashort pulses in a microresonator and offer broadband and coherent optical frequency combs \cite{Herr:14}. A number of applications range from coherent communications \cite{Marin-Palomo:17}, dual-comb spectroscopy \cite{Suh:16}, astrophysical spectrometer calibration \cite{Obrzud:19, Suh:19c}, ultrafast distance measurement \cite{Suh:18,Trocha:18} to ultra-low noise microwave generation \cite{Liang:15, Liu:19} have been demonstrated utilizing DKS. Platforms that generated DKS include $\mathrm {MgF_2}$, $\mathrm {SiO_2}$,  $\mathrm {Si_3N_4}$, Si, AlN, and $\mathrm {LiNbO_3}$ \cite{Kippenberg:18}.  One of the challenges to realize a fully-integrated soliton microcomb was to obtain a high $Q$-factor in photonic integrated microresonator ($e.g.$ $\mathrm {Si_3N_4}$) in order to avoid the use of an external amplifier to meet  power requirements. Moreover, the complex soliton tuning mechanism typically requires lasers with stabilized modules and additional actuators to overcome the thermal effects due to operation at high power \cite{Brasch:16, Stone:18}. In parallel, a  soliton  microcomb based on ultrahigh-$Q$  silica air-clad microresonator  integrated with  $\rm Si_3N_4$ waveguide have been shown\cite{Yang:18}. Recently,   current-initiated soliton microcombs based on an integrated $\rm Si_3N_4$ microresonator have been demonstrated, using injection locking to either a multimode laser diode, or to a cavity formed with the photonic chip. Both approaches succeeded due to significant reductions in optical losses \cite{Stern:18, Raja:19}. Still, these approaches suffer from limited input power levels which limit the comb tooth power due to low nonlinear conversion efficiency. Moreover, electrically-driven \emph{single} solitons have only been demonstrated at a repetition rate of 149 GHz using this approach \cite{Raja:19}, which is difficult to detect electronically.  A semiconductor-based fully packaged and fiber-coupled hyrid laser operating at relatively high power ($\sim $100 mW), and combing low relative intensity noise (–160 dBc/Hz at $\mathrm f_\mathrm{offset}$ $\sim$ 100 kHz) with narrow linewidth (as low as $15 \mathrm{Hz}$) is correspondingly a promising candidate for generating soliton microcombs \cite{Morton:18}. By packaging the $\mathrm{Si_3N_4}$ microresonator, a very compact soliton microcomb system can be demonstrated which may enable the use of such devices in practical applications. Recently, a soliton microcomb operating in the microwave K- and X-band was demonstrated using photonic integrated  $\mathrm{Si_3N_4}$ microresonators at a very low input power ($\sim$ 35 mW on-chip) \cite{Liu:19}. As pointed out in that work, given that the laser noise is transferred to the soliton comb,  it is critical to use a laser with low frequency noise  that can be operated at  sufficiently high power ($>$ 60 mW) to initiate soliton generation.

Here, we demonstrate an ultra-compact photonic integrated soliton microcomb of 99 GHz repetition rate by using a commercially available ultra-compact, low-noise and high-power hybrid laser \cite{Morton:18}. The $\mathrm{Si_3N_4}$ microresonators are fabricated using the photonic Damascene reflow process enabling ultra-smooth waveguide sidewalls and intrinsic $Q$-factors exceeding  $1.5\times10^7$ across the telecom C- and L-band \cite{Pfeiffer:18, Liu:18a}.  The soliton microcomb is initiated via tuning of the laser diode current, which reduces significantly the system complexity. Different comb and soliton states are accessed deterministically by adjusting the current of the laser diode \citep{Herr:14}. A heterodyne beatnote measurement is performed to verify the coherence of the soliton state. Furthermore, a packaging technique is developed for achieving a compact, portable microcomb system. A similar approach of direct pumping also has been shown using a silica microresonator to generate soliton microcomb \cite{Nick:18,Suh:19}.


 
  

\begin{figure}[t!]
\centering
\includegraphics[width=1 \linewidth]{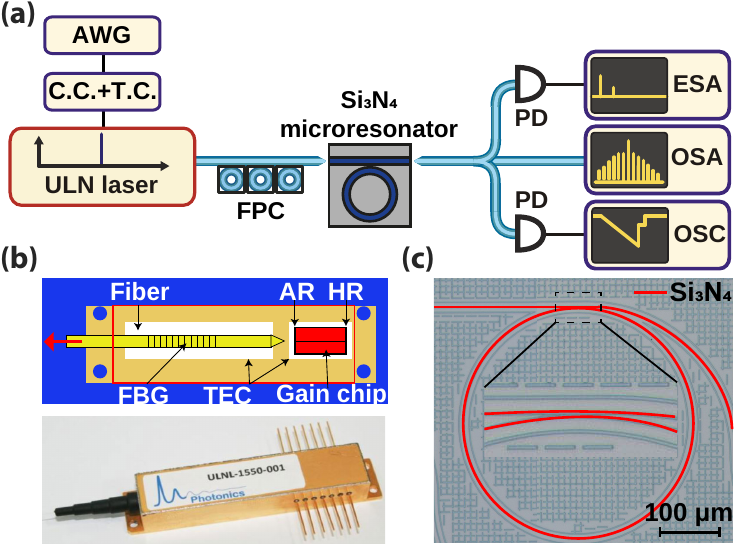}
\caption{\textbf{Experimental setup for the generation of a 99-GHz soliton microcomb using the hybrid ULN laser.} \textbf{a)} The compact hybrid laser consisting of a gain chip (GC) and a fiber Bragg grating (FBG) is coupled to a  $\mathrm{Si_3N_4}$ microresonator chip. The laser is current- (C.C.) and temperature-controlled (T.C.). An arbitrary waveform generator (AWG) is used to scan over the resonance by directly modulating the laser current. The generated soliton microcombs are analysed in the optical and electrical domain using an optical spectrum analyser (OSA) and an electrical spectrum analyser (ESA) respectively. OSC: Oscilloscope and FPC: Fiber polarization controller. \textbf{b)} A butterfly packaged ultra-low noise and high power laser (bottom panel). The ULN laser consists of a gain chip and an external fiber cavity (top panel). AR: Anti reflective, HR: Highly reflective, and TEC: Thermoelectric cooler  \textbf{c)} A  $\mathrm{Si_3N_4}$ photonic chip based  microresonator.  }
\label{Fig:Fig1}
\end{figure}

The ultralow-noise (ULN) laser consists of two parts (Fig. \ref{Fig:Fig1}b): The first part is a semiconductor based gain chip facilitating high power operation.  It features a highly reflective facet on one side and an angled facet on the other side for out-coupling \cite{Morton:18}. The second part is a customized fiber Bragg grating (FBG) that supports single frequency and narrow linewidth operation. The light from the gain chip  is coupled efficiently to the FBG via polarization maintaining (PM) lensed fiber. The laser power and frequency can be tuned via current control applied on the gain chip, and temperature control applied on the gain chip or the FBG. The maximum output power is around 100 mW. The laser center frequency is approximately 193.4 THz and has a tuning range of ${\sim}$ 100 GHz. An isolator is used at output to avoid any back reflection into the laser. 

\begin{figure}[htbp]
\centering
\includegraphics[]{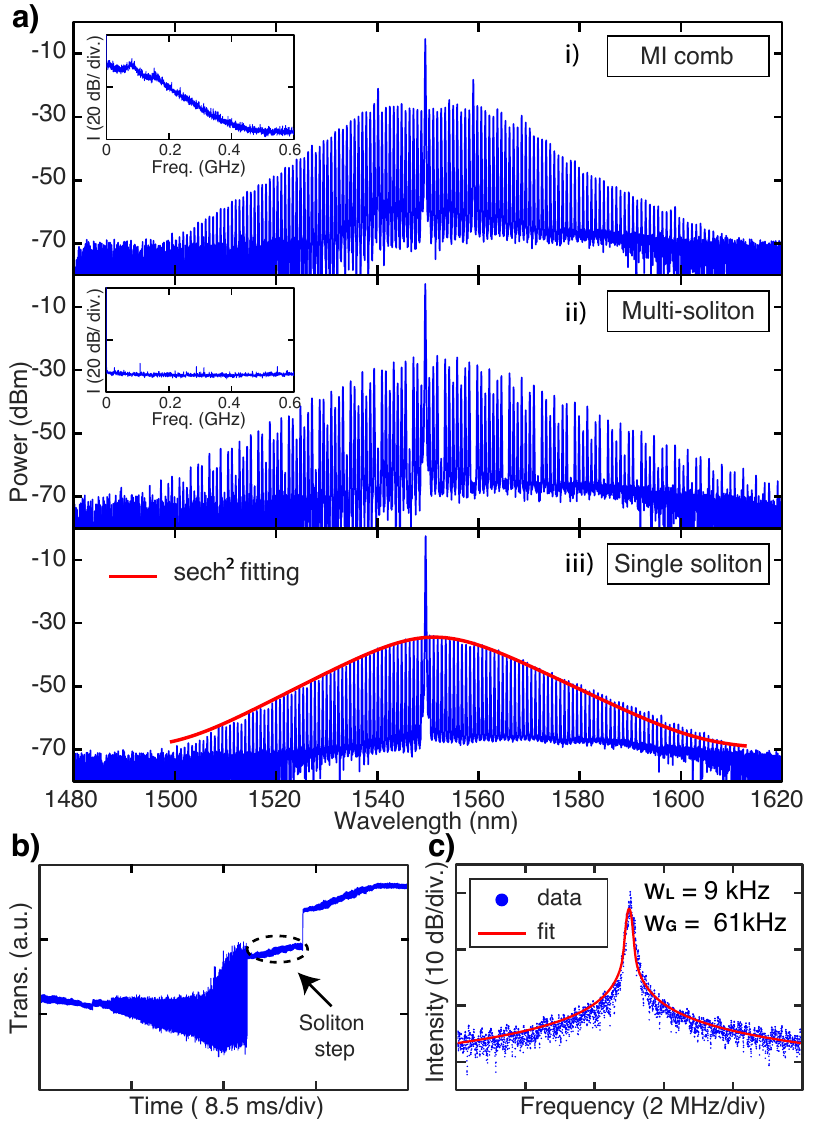}
\caption{\textbf{Different microcomb states  generated in a 99-GHz-FSR $\mathrm{Si_3N_4}$ microresonator  using a compact ULN laser}. \textbf{a)} A modulation instability state generated by modulating the current applied to laser diode (i). By adjusting the end point of the scan a low-noise multi-soliton state is generated (ii). A low radio-frequency (RF) spectrum  is recorded to access coherence properties of  the different comb states (inset). A single soliton state with a characteristic $\mathrm {sech^2(\textit{f})}$ profile, featuring a 3-dB bandwidth of 19 nm and a pulse duration of 131-fs (iii). \textbf{b)} Transmission signal detected on the photodetector showing a soliton step. The soliton is initiated by scanning over the microresonator resonance from the effective blue-detuned side to the red-detuned side. \textbf{c)} Heterodyne beatnote signal fitted with a Voigt profile is showing a Lorentzian  linewidth of 9 kHz and a Gaussian linewidth of 61 kHz. }
\label{Fig:Fig4}
\end{figure}

\begin{figure*}[htbp]
\centering
\includegraphics[width=1 \linewidth]{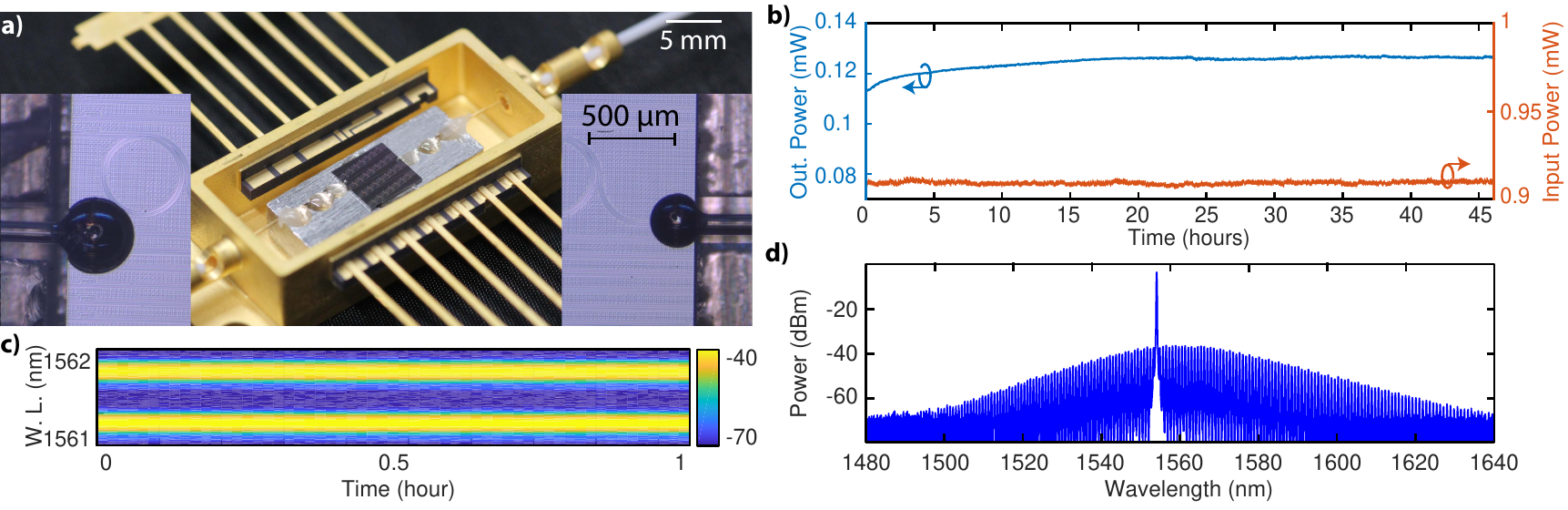}
\caption{\textbf{A  $\mathrm{Si_3N_4}$ microcomb system in a compact package}. \textbf{a)} Photonic package of a $\mathrm{Si_3N_4}$ chip using  an  ultrahigh-numerical-aperture (UHNA)  fiber spliced with a  SMF-28 inside a butterfly package. Insets: zoomed-in view of the fiber-chip facet showing the drop of the epoxy used to attach the fiber. A low-shrinkage and medium viscous epoxy has been used to minimize losses (EPO-TEK OG154). \textbf{b)} Long-term monitoring of the coupling of the packaged system to evaluate the performance of the device. Initially 5 to 7 hours, the slow increase in output power can be associated with curing of epoxy. The epoxy usually takes 24 hours to settle down after  UV curing. \textbf{c)} A long-term stability test in the packaged system showing two lines of a single soliton microcomb (spectrum in \textbf{d}). The soliton state was maintained for more than one hour without any active chip temperature stabilization (fully free running). W. L.: Wavelength. \textbf{d)} Single soliton generation using a different laser diode in a packaged device.    }
\label{Fig:Fig6}
\end{figure*}
The $\rm Si_3N_4$ microresonator chip is fabricated using the photonic Damascene reflow process \cite{Pfeiffer:18, Pfeiffer:18a}. The microresonator and bus waveguide are both 1.5 $\mathrm{\mu m}$ wide and 700 nm high to achieve high coupling ideality \cite{Pfeiffer:17b}. The light is coupled to the $\rm Si_3N_4$ bus waveguide via double-inverse tapers (width $ \sim$ 300 nm) on chip facets (in- and out-coupling) \cite{Liu:18}, with $>25\%$ coupling efficiency (fiber-chip-fiber). A polarization controller is used to align the polarization of incoming light to the transverse-electric (TE) mode.  The $\mathrm {Si_3N_4}$ photonic chips have been characterized using a diode laser spectroscopy technique, the calibration is performed with a fully stabilized commercial optical frequency comb system \cite{DelHaye:09, Liu:16}. Every resonance has been fitted using the model explained in Ref.~\cite{Gorodetsky:00} to extract the intrinsic linewidth $\kappa_0 / 2\pi$, coupling strength $\kappa_{\mathrm{ex}} / 2\pi$, and backscattering rate $\gamma / 2 \pi$. The microresonators show a  most probable value of  intrinsic $Q$-factor of $Q_0\sim 1.5\times10^7$ across the telecom C- and L-band \cite{Liu:18a}. Single soliton states with 99 GHz repetition rate have  already been demonstrated in these high $Q$-factor $\mathrm{Si_3N_4}$ microresonators, using only a diode laser at a record-low output power of 9 mW \cite{Liu:18a}. However, that scheme required a well stabilized  bulk laser diode (Toptica CTL 1550) which is not feasible in a compact package and not compatible with further potential photonic integration.


The hybrid ULN laser is directly coupled to the $\rm Si_3N_4$ chip without an additional amplifier, as shown in Fig. \ref{Fig:Fig1}. First, the laser is operated at a few milliwatt of output power and the laser frequency is tuned to a microresonator resonance by observing the transmission signal detected on the photodiode. The temperature of the FBG is adjusted in small steps until a Lorentzian signal is observed on the oscilloscope. The laser power is further increased by acting on the current ($\sim$ 300 mA ) applied on the laser diode, until a step-like pattern is observed in the transmission signal, signalling  soliton formation (Fig. \ref{Fig:Fig4}b). As the laser frequency changes simultaneously, the temperature of FBG must be adjusted accordingly to remain on resonance. 
In order to overcome the thermal effects in microresonators, complex soliton tuning techniques have been reported, such as using power kicking \cite{Brasch:16}, single sideband modulator \cite{Stone:18} and dual lasers \cite{Zhang:19}, which  require off-chip components and can complicate the photonic packaging and integration on $\rm Si_3N_4$ chips. 
Here we use only the forward laser frequency tuning method \cite{Herr:14} to access the soliton state, in which conventionally a triangular shape voltage signal is applied to the laser piezo to scan over the resonance. However, the fast tuning signal cannot be applied due to limited piezo scan speed. Here, we directly modulate the current applied to the laser diode, allowing fast scan over resonance ($\mathrm{\sim kHz}$), which in turn modulates the laser frequency and power to generate different comb states without any complexity. The soliton existence range is sufficiently high (Fig. \ref{Fig:Fig4}b, inset) due to the high $Q$-factor and the reduced thermal effects obtained from a novel annealing scheme \cite{Liu:18a}, thus simultaneous modulation of laser frequency and power does not affect while tuning into the soliton state. The reason being that, as the soliton exists on the effectively red detuned side, the increase of current applied to the laser diode leads to an increase in the wavelength as well as in optical power. The increase in input power enhances the soliton existence range. A baseband radio-frequency (RF) spectrum is recorded to verify the coherence of the different comb states (Fig. \ref{Fig:Fig4}a, inset). A modulation instability (MI) comb state was generated initially by adjusting the end point of the voltage signal applied to modulate the current. Afterwards, a multiple soliton state is initiated by increasing the current and performing a scan over resonance from the blue-detuned to the red-detuned side. A low intensity noise  is observed on the electrical spectrum analyser (ESA), indicating the coherent nature of the soliton state. Once all  parameters such as current applied to laser diode and temperatures (GC and FBG)  are known, a soliton can be generated deterministically  by performing a scan over the resonance, which  simplifies the soliton initiation process. Heterodyne beatnote measurements are carried out to further confirm the intrinsic coherence of the soliton state. A narrow-linewidth ($\sim$ 10 kHz) reference laser is used to generate a beatnote with a tooth of the soliton comb. The measured beatnote is fitted with a Voigt profile and shows a Lorentzian linewidth of 9 kHz and a Gaussian linewidth of 61 kHz.  The  linewidth is likely limited by the current fluctuations of the laser driver (Fig. \ref{Fig:Fig4}c). The single soliton spectrum has been fitted with an hyperbolic secant squared ($\mathrm{sech^2} $) function. The fitting shows a a pulse duration of 131-fs and a 3-dB bandwidth of 19 nm (Fig.\ref{Fig:Fig4}a, iii).


Next, we demonstrate a compact soliton microcomb module by robustly packaging  the $\mathrm{Si_3N_4}$ chip. A 2-cm-long ultrahigh-numerical-aperture (UHNA) fiber (mode field diameter $\sim$ 4.1 $\mu$m) is used to mode-match the fiber mode to the taper waveguide mode on the $\mathrm{Si_3N_4}$ chips, and is spliced with an SMF-28 fiber ($\sim$ 1 dB splicing loss). Fiber-chip-fiber through coupling efficiency of 15\% is achieved using the UHNA fiber (including the splicing loss). The splicing loss is minimized by performing a multiple arc discharge which allows the UHNA fiber core to expand for adiabatic mode conversion. The fiber is aligned actively to the chip input facet mechanically, and then a drop of  epoxy is dispensed on the fiber-chip interface using an accurate ($\sim$ 20 $\mathrm{\mu m}$) pneumatic valve. One important aspect of the packaging is to keep the initial drop size as small as possible (Fig. \ref{Fig:Fig6}a, inset), to avoid extra loss due to dimensional  changes which occur during the curing of the glue. After dispensing the glue, UV curing is performed in 3 steps with different UV light intensity for optimal curing performance: 100 mW/cm$^2$ for 0.5 - 1 min, 200 mW/cm$^2$ for 1 - 3 min and 300 mW/cm$^2$ for 3 - 5 min. After  UV curing, the packaged device is tested for long-term stability with input laser power below 1 mW. The coupling remains stable for more than 30 hours, showing  the robustness of the packaged device (Fig. \ref{Fig:Fig6}b). Afterwards, the input power is increased in order to generate a microcomb in the $\mathrm{Si_3N_4}$ chip. Due to the higher coupling loss as compared to the case using lensed fibers, only MI comb states can be generated in the packaged device while pumping with the ULN laser. The coupling losses can be reduced by packaging with lensed fibers or improving the on-chip ($\mathrm{Si_3N_4}$) mode converter  to generate the soliton states using a compact laser. To demonstrate the stability of the packaged device at the high power, a different diode laser along with EDFA is used to generate a single soliton in a 99-GHz microresonator at an input power of $\sim$  150 mW. The packaged device can be operated in a single soliton  state for more than one hour without any active stabilization, indicating that such device can be field-deployable (Fig. \ref{Fig:Fig6}c). 

In summary, we have demonstrated a soliton microcomb operating at 99 GHz repetition rate directly driven by a compact laser. The single soliton state is initiated exclusively via current tuning of the laser diode. A packaging technique has been developed for a compact, portable microcomb system. Such a compact system is  crucial for  applications requiring a high power per comb line and low phase noise, such as coherent communications \cite{Marin-Palomo:17} and on-chip low-noise microwave generation \cite{Liu:19}.

\section*{Funding Information}
This publication was supported by Contract HR0011-15-C-0055 from the Defense Advanced Research Projects Agency (DARPA) Microsystems Technology Office (MTO), by the Air Force Office of Scientific Research, Air Force Material Command, USAF under Award No. FA9550-15-1-0099, by Swiss National Science Foundation (SNSF) (176563, BRIDGE), and by FET grant agreement No. 801352 (OPTIMISM). E.L.  acknowledges the support from the European Space Technology
Centre with ESA Contract No. 4000116145/16/NL/MH/GM. J.H. acknowledges the support provided by Prof. Hwa-Yaw Tam and from the General Research Fund of the Hong Kong Government under project PolyU152207/15E.

\section*{Acknowledgments}
The $\mathbf{Si_3N_4}$ microresonator samples were fabricated in the EPFL center of Micro-NanoTechnology (CMi).  We thank J. C. Skehan and J. Wachs for useful discussion and  A. Lukashchuk for assistance in chip mask design.

\noindent \textbf{Data Availability Statement}: The code and data used to produce the plots within this work will be released on the repository \texttt{Zenodo} upon publication of this preprint.

\bibliographystyle{apsrev4-1}
\bibliography{bibliography}

\begin{thebibliography}{29}%
\makeatletter
\providecommand \@ifxundefined [1]{%
 \@ifx{#1\undefined}
}%
\providecommand \@ifnum [1]{%
 \ifnum #1\expandafter \@firstoftwo
 \else \expandafter \@secondoftwo
 \fi
}%
\providecommand \@ifx [1]{%
 \ifx #1\expandafter \@firstoftwo
 \else \expandafter \@secondoftwo
 \fi
}%
\providecommand \natexlab [1]{#1}%
\providecommand \enquote  [1]{``#1''}%
\providecommand \bibnamefont  [1]{#1}%
\providecommand \bibfnamefont [1]{#1}%
\providecommand \citenamefont [1]{#1}%
\providecommand \href@noop [0]{\@secondoftwo}%
\providecommand \href [0]{\begingroup \@sanitize@url \@href}%
\providecommand \@href[1]{\@@startlink{#1}\@@href}%
\providecommand \@@href[1]{\endgroup#1\@@endlink}%
\providecommand \@sanitize@url [0]{\catcode `\\12\catcode `\$12\catcode
  `\&12\catcode `\#12\catcode `\^12\catcode `\_12\catcode `\%12\relax}%
\providecommand \@@startlink[1]{}%
\providecommand \@@endlink[0]{}%
\providecommand \url  [0]{\begingroup\@sanitize@url \@url }%
\providecommand \@url [1]{\endgroup\@href {#1}{\urlprefix }}%
\providecommand \urlprefix  [0]{URL }%
\providecommand \Eprint [0]{\href }%
\providecommand \doibase [0]{http://dx.doi.org/}%
\providecommand \selectlanguage [0]{\@gobble}%
\providecommand \bibinfo  [0]{\@secondoftwo}%
\providecommand \bibfield  [0]{\@secondoftwo}%
\providecommand \translation [1]{[#1]}%
\providecommand \BibitemOpen [0]{}%
\providecommand \bibitemStop [0]{}%
\providecommand \bibitemNoStop [0]{.\EOS\space}%
\providecommand \EOS [0]{\spacefactor3000\relax}%
\providecommand \BibitemShut  [1]{\csname bibitem#1\endcsname}%
\let\auto@bib@innerbib\@empty
\bibitem [{\citenamefont {Udem}\ \emph {et~al.}(2002)\citenamefont {Udem},
  \citenamefont {Holzwarth},\ and\ \citenamefont {H{\"a}nsch}}]{Udem:02}%
  \BibitemOpen
  \bibfield  {author} {\bibinfo {author} {\bibfnamefont {T.}~\bibnamefont
  {Udem}}, \bibinfo {author} {\bibfnamefont {R.}~\bibnamefont {Holzwarth}}, \
  and\ \bibinfo {author} {\bibfnamefont {T.~W.}\ \bibnamefont {H{\"a}nsch}},\
  }\href {http://dx.doi.org/10.1038/416233a} {\bibfield  {journal} {\bibinfo
  {journal} {Nature}\ }\textbf {\bibinfo {volume} {416}},\ \bibinfo {pages}
  {233} (\bibinfo {year} {2002})}\BibitemShut {NoStop}%
\bibitem [{\citenamefont {Del'Haye}\ \emph {et~al.}(2007)\citenamefont
  {Del'Haye}, \citenamefont {Schliesser}, \citenamefont {Arcizet},
  \citenamefont {Wilken}, \citenamefont {Holzwarth},\ and\ \citenamefont
  {Kippenberg}}]{DelHaye:07}%
  \BibitemOpen
  \bibfield  {author} {\bibinfo {author} {\bibfnamefont {P.}~\bibnamefont
  {Del'Haye}}, \bibinfo {author} {\bibfnamefont {A.}~\bibnamefont
  {Schliesser}}, \bibinfo {author} {\bibfnamefont {O.}~\bibnamefont {Arcizet}},
  \bibinfo {author} {\bibfnamefont {T.}~\bibnamefont {Wilken}}, \bibinfo
  {author} {\bibfnamefont {R.}~\bibnamefont {Holzwarth}}, \ and\ \bibinfo
  {author} {\bibfnamefont {T.~J.}\ \bibnamefont {Kippenberg}},\ }\href@noop {}
  {\bibfield  {journal} {\bibinfo  {journal} {Nature}\ }\textbf {\bibinfo
  {volume} {450}},\ \bibinfo {pages} {1214} (\bibinfo {year}
  {2007})}\BibitemShut {NoStop}%
\bibitem [{\citenamefont {Kippenberg}\ \emph {et~al.}(2018)\citenamefont
  {Kippenberg}, \citenamefont {Gaeta}, \citenamefont {Lipson},\ and\
  \citenamefont {Gorodetsky}}]{Kippenberg:18}%
  \BibitemOpen
  \bibfield  {author} {\bibinfo {author} {\bibfnamefont {T.~J.}\ \bibnamefont
  {Kippenberg}}, \bibinfo {author} {\bibfnamefont {A.~L.}\ \bibnamefont
  {Gaeta}}, \bibinfo {author} {\bibfnamefont {M.}~\bibnamefont {Lipson}}, \
  and\ \bibinfo {author} {\bibfnamefont {M.~L.}\ \bibnamefont {Gorodetsky}},\
  }\href {\doibase 10.1126/science.aan8083} {\bibfield  {journal} {\bibinfo
  {journal} {Science}\ }\textbf {\bibinfo {volume} {361}} (\bibinfo {year}
  {2018}),\ 10.1126/science.aan8083}\BibitemShut {NoStop}%
\bibitem [{\citenamefont {Herr}\ \emph {et~al.}(2013)\citenamefont {Herr},
  \citenamefont {Brasch}, \citenamefont {Jost}, \citenamefont {Wang},
  \citenamefont {Kondratiev}, \citenamefont {Gorodetsky},\ and\ \citenamefont
  {Kippenberg}}]{Herr:14}%
  \BibitemOpen
  \bibfield  {author} {\bibinfo {author} {\bibfnamefont {T.}~\bibnamefont
  {Herr}}, \bibinfo {author} {\bibfnamefont {V.}~\bibnamefont {Brasch}},
  \bibinfo {author} {\bibfnamefont {J.~D.}\ \bibnamefont {Jost}}, \bibinfo
  {author} {\bibfnamefont {C.~Y.}\ \bibnamefont {Wang}}, \bibinfo {author}
  {\bibfnamefont {N.~M.}\ \bibnamefont {Kondratiev}}, \bibinfo {author}
  {\bibfnamefont {M.~L.}\ \bibnamefont {Gorodetsky}}, \ and\ \bibinfo {author}
  {\bibfnamefont {T.~J.}\ \bibnamefont {Kippenberg}},\ }\href
  {https://doi.org/10.1038/nphoton.2013.343} {\bibfield  {journal} {\bibinfo
  {journal} {Nature Photonics}\ }\textbf {\bibinfo {volume} {8}},\ \bibinfo
  {pages} {145 EP } (\bibinfo {year} {2013})}\BibitemShut {NoStop}%
\bibitem [{\citenamefont {Marin-Palomo}\ \emph {et~al.}(2017)\citenamefont
  {Marin-Palomo}, \citenamefont {Kemal}, \citenamefont {Karpov}, \citenamefont
  {Kordts}, \citenamefont {Pfeifle}, \citenamefont {Pfeiffer}, \citenamefont
  {Trocha}, \citenamefont {Wolf}, \citenamefont {Brasch}, \citenamefont
  {Anderson}, \citenamefont {Rosenberger}, \citenamefont {Vijayan},
  \citenamefont {Freude}, \citenamefont {Kippenberg},\ and\ \citenamefont
  {Koos}}]{Marin-Palomo:17}%
  \BibitemOpen
  \bibfield  {author} {\bibinfo {author} {\bibfnamefont {P.}~\bibnamefont
  {Marin-Palomo}}, \bibinfo {author} {\bibfnamefont {J.~N.}\ \bibnamefont
  {Kemal}}, \bibinfo {author} {\bibfnamefont {M.}~\bibnamefont {Karpov}},
  \bibinfo {author} {\bibfnamefont {A.}~\bibnamefont {Kordts}}, \bibinfo
  {author} {\bibfnamefont {J.}~\bibnamefont {Pfeifle}}, \bibinfo {author}
  {\bibfnamefont {M.~H.~P.}\ \bibnamefont {Pfeiffer}}, \bibinfo {author}
  {\bibfnamefont {P.}~\bibnamefont {Trocha}}, \bibinfo {author} {\bibfnamefont
  {S.}~\bibnamefont {Wolf}}, \bibinfo {author} {\bibfnamefont {V.}~\bibnamefont
  {Brasch}}, \bibinfo {author} {\bibfnamefont {M.~H.}\ \bibnamefont
  {Anderson}}, \bibinfo {author} {\bibfnamefont {R.}~\bibnamefont
  {Rosenberger}}, \bibinfo {author} {\bibfnamefont {K.}~\bibnamefont
  {Vijayan}}, \bibinfo {author} {\bibfnamefont {W.}~\bibnamefont {Freude}},
  \bibinfo {author} {\bibfnamefont {T.~J.}\ \bibnamefont {Kippenberg}}, \ and\
  \bibinfo {author} {\bibfnamefont {C.}~\bibnamefont {Koos}},\ }\href
  {https://doi.org/10.1038/nature22387} {\bibfield  {journal} {\bibinfo
  {journal} {Nature}\ }\textbf {\bibinfo {volume} {546}},\ \bibinfo {pages}
  {274 EP } (\bibinfo {year} {2017})}\BibitemShut {NoStop}%
\bibitem [{\citenamefont {Suh}\ \emph {et~al.}(2016)\citenamefont {Suh},
  \citenamefont {Yang}, \citenamefont {Yang}, \citenamefont {Yi},\ and\
  \citenamefont {Vahala}}]{Suh:16}%
  \BibitemOpen
  \bibfield  {author} {\bibinfo {author} {\bibfnamefont {M.-G.}\ \bibnamefont
  {Suh}}, \bibinfo {author} {\bibfnamefont {Q.-F.}\ \bibnamefont {Yang}},
  \bibinfo {author} {\bibfnamefont {K.~Y.}\ \bibnamefont {Yang}}, \bibinfo
  {author} {\bibfnamefont {X.}~\bibnamefont {Yi}}, \ and\ \bibinfo {author}
  {\bibfnamefont {K.~J.}\ \bibnamefont {Vahala}},\ }\href {\doibase
  10.1126/science.aah6516} {\bibfield  {journal} {\bibinfo  {journal}
  {Science}\ }\textbf {\bibinfo {volume} {354}},\ \bibinfo {pages} {600}
  (\bibinfo {year} {2016})}\BibitemShut {NoStop}%
\bibitem [{\citenamefont {Obrzud}\ \emph {et~al.}(2019)\citenamefont {Obrzud},
  \citenamefont {Rainer}, \citenamefont {Harutyunyan}, \citenamefont
  {Anderson}, \citenamefont {Liu}, \citenamefont {Geiselmann}, \citenamefont
  {Chazelas}, \citenamefont {Kundermann}, \citenamefont {Lecomte},
  \citenamefont {Cecconi}, \citenamefont {Ghedina}, \citenamefont {Molinari},
  \citenamefont {Pepe}, \citenamefont {Wildi}, \citenamefont {Bouchy},
  \citenamefont {Kippenberg},\ and\ \citenamefont {Herr}}]{Obrzud:19}%
  \BibitemOpen
  \bibfield  {author} {\bibinfo {author} {\bibfnamefont {E.}~\bibnamefont
  {Obrzud}}, \bibinfo {author} {\bibfnamefont {M.}~\bibnamefont {Rainer}},
  \bibinfo {author} {\bibfnamefont {A.}~\bibnamefont {Harutyunyan}}, \bibinfo
  {author} {\bibfnamefont {M.~H.}\ \bibnamefont {Anderson}}, \bibinfo {author}
  {\bibfnamefont {J.}~\bibnamefont {Liu}}, \bibinfo {author} {\bibfnamefont
  {M.}~\bibnamefont {Geiselmann}}, \bibinfo {author} {\bibfnamefont
  {B.}~\bibnamefont {Chazelas}}, \bibinfo {author} {\bibfnamefont
  {S.}~\bibnamefont {Kundermann}}, \bibinfo {author} {\bibfnamefont
  {S.}~\bibnamefont {Lecomte}}, \bibinfo {author} {\bibfnamefont
  {M.}~\bibnamefont {Cecconi}}, \bibinfo {author} {\bibfnamefont
  {A.}~\bibnamefont {Ghedina}}, \bibinfo {author} {\bibfnamefont
  {E.}~\bibnamefont {Molinari}}, \bibinfo {author} {\bibfnamefont
  {F.}~\bibnamefont {Pepe}}, \bibinfo {author} {\bibfnamefont {F.}~\bibnamefont
  {Wildi}}, \bibinfo {author} {\bibfnamefont {F.}~\bibnamefont {Bouchy}},
  \bibinfo {author} {\bibfnamefont {T.~J.}\ \bibnamefont {Kippenberg}}, \ and\
  \bibinfo {author} {\bibfnamefont {T.}~\bibnamefont {Herr}},\ }\href {\doibase
  10.1038/s41566-018-0309-y} {\bibfield  {journal} {\bibinfo  {journal} {Nature
  Photonics}\ }\textbf {\bibinfo {volume} {13}},\ \bibinfo {pages} {31}
  (\bibinfo {year} {2019})}\BibitemShut {NoStop}%
\bibitem [{\citenamefont {Suh}\ \emph {et~al.}(2019{\natexlab{a}})\citenamefont
  {Suh}, \citenamefont {Yi}, \citenamefont {Lai}, \citenamefont {Leifer},
  \citenamefont {Grudinin}, \citenamefont {Vasisht}, \citenamefont {Martin},
  \citenamefont {Fitzgerald}, \citenamefont {Doppmann}, \citenamefont {Wang},
  \citenamefont {Mawet}, \citenamefont {Papp}, \citenamefont {Diddams},
  \citenamefont {Beichman},\ and\ \citenamefont {Vahala}}]{Suh:19c}%
  \BibitemOpen
  \bibfield  {author} {\bibinfo {author} {\bibfnamefont {M.-G.}\ \bibnamefont
  {Suh}}, \bibinfo {author} {\bibfnamefont {X.}~\bibnamefont {Yi}}, \bibinfo
  {author} {\bibfnamefont {Y.-H.}\ \bibnamefont {Lai}}, \bibinfo {author}
  {\bibfnamefont {S.}~\bibnamefont {Leifer}}, \bibinfo {author} {\bibfnamefont
  {I.~S.}\ \bibnamefont {Grudinin}}, \bibinfo {author} {\bibfnamefont
  {G.}~\bibnamefont {Vasisht}}, \bibinfo {author} {\bibfnamefont {E.~C.}\
  \bibnamefont {Martin}}, \bibinfo {author} {\bibfnamefont {M.~P.}\
  \bibnamefont {Fitzgerald}}, \bibinfo {author} {\bibfnamefont
  {G.}~\bibnamefont {Doppmann}}, \bibinfo {author} {\bibfnamefont
  {J.}~\bibnamefont {Wang}}, \bibinfo {author} {\bibfnamefont {D.}~\bibnamefont
  {Mawet}}, \bibinfo {author} {\bibfnamefont {S.~B.}\ \bibnamefont {Papp}},
  \bibinfo {author} {\bibfnamefont {S.~A.}\ \bibnamefont {Diddams}}, \bibinfo
  {author} {\bibfnamefont {C.}~\bibnamefont {Beichman}}, \ and\ \bibinfo
  {author} {\bibfnamefont {K.}~\bibnamefont {Vahala}},\ }\href {\doibase
  10.1038/s41566-018-0312-3} {\bibfield  {journal} {\bibinfo  {journal} {Nature
  Photonics}\ }\textbf {\bibinfo {volume} {13}},\ \bibinfo {pages} {25}
  (\bibinfo {year} {2019}{\natexlab{a}})}\BibitemShut {NoStop}%
\bibitem [{\citenamefont {Suh}\ and\ \citenamefont {Vahala}(2018)}]{Suh:18}%
  \BibitemOpen
  \bibfield  {author} {\bibinfo {author} {\bibfnamefont {M.-G.}\ \bibnamefont
  {Suh}}\ and\ \bibinfo {author} {\bibfnamefont {K.~J.}\ \bibnamefont
  {Vahala}},\ }\href {\doibase 10.1126/science.aao1968} {\bibfield  {journal}
  {\bibinfo  {journal} {Science}\ }\textbf {\bibinfo {volume} {359}},\ \bibinfo
  {pages} {884} (\bibinfo {year} {2018})}\BibitemShut {NoStop}%
\bibitem [{\citenamefont {Trocha}\ \emph {et~al.}(2018)\citenamefont {Trocha},
  \citenamefont {Karpov}, \citenamefont {Ganin}, \citenamefont {Pfeiffer},
  \citenamefont {Kordts}, \citenamefont {Wolf}, \citenamefont {Krockenberger},
  \citenamefont {Marin-Palomo}, \citenamefont {Weimann}, \citenamefont
  {Randel}, \citenamefont {Freude}, \citenamefont {Kippenberg},\ and\
  \citenamefont {Koos}}]{Trocha:18}%
  \BibitemOpen
  \bibfield  {author} {\bibinfo {author} {\bibfnamefont {P.}~\bibnamefont
  {Trocha}}, \bibinfo {author} {\bibfnamefont {M.}~\bibnamefont {Karpov}},
  \bibinfo {author} {\bibfnamefont {D.}~\bibnamefont {Ganin}}, \bibinfo
  {author} {\bibfnamefont {M.~H.~P.}\ \bibnamefont {Pfeiffer}}, \bibinfo
  {author} {\bibfnamefont {A.}~\bibnamefont {Kordts}}, \bibinfo {author}
  {\bibfnamefont {S.}~\bibnamefont {Wolf}}, \bibinfo {author} {\bibfnamefont
  {J.}~\bibnamefont {Krockenberger}}, \bibinfo {author} {\bibfnamefont
  {P.}~\bibnamefont {Marin-Palomo}}, \bibinfo {author} {\bibfnamefont
  {C.}~\bibnamefont {Weimann}}, \bibinfo {author} {\bibfnamefont
  {S.}~\bibnamefont {Randel}}, \bibinfo {author} {\bibfnamefont
  {W.}~\bibnamefont {Freude}}, \bibinfo {author} {\bibfnamefont {T.~J.}\
  \bibnamefont {Kippenberg}}, \ and\ \bibinfo {author} {\bibfnamefont
  {C.}~\bibnamefont {Koos}},\ }\href {\doibase 10.1126/science.aao3924}
  {\bibfield  {journal} {\bibinfo  {journal} {Science}\ }\textbf {\bibinfo
  {volume} {359}},\ \bibinfo {pages} {887} (\bibinfo {year}
  {2018})}\BibitemShut {NoStop}%
\bibitem [{\citenamefont {Liang}\ \emph {et~al.}(2015)\citenamefont {Liang},
  \citenamefont {Eliyahu}, \citenamefont {Ilchenko}, \citenamefont
  {Savchenkov}, \citenamefont {Matsko}, \citenamefont {Seidel},\ and\
  \citenamefont {Maleki}}]{Liang:15}%
  \BibitemOpen
  \bibfield  {author} {\bibinfo {author} {\bibfnamefont {W.}~\bibnamefont
  {Liang}}, \bibinfo {author} {\bibfnamefont {D.}~\bibnamefont {Eliyahu}},
  \bibinfo {author} {\bibfnamefont {V.~S.}\ \bibnamefont {Ilchenko}}, \bibinfo
  {author} {\bibfnamefont {A.~A.}\ \bibnamefont {Savchenkov}}, \bibinfo
  {author} {\bibfnamefont {A.~B.}\ \bibnamefont {Matsko}}, \bibinfo {author}
  {\bibfnamefont {D.}~\bibnamefont {Seidel}}, \ and\ \bibinfo {author}
  {\bibfnamefont {L.}~\bibnamefont {Maleki}},\ }\href
  {https://doi.org/10.1038/ncomms8957} {\bibfield  {journal} {\bibinfo
  {journal} {Nature Communications}\ }\textbf {\bibinfo {volume} {6}},\
  \bibinfo {pages} {7957} (\bibinfo {year} {2015})}\BibitemShut {NoStop}%
\bibitem [{\citenamefont {Liu}\ \emph {et~al.}(2019)\citenamefont {Liu},
  \citenamefont {Lucas}, \citenamefont {He}, \citenamefont {Raja},
  \citenamefont {Wang}, \citenamefont {Karpov}, \citenamefont {Guo},
  \citenamefont {Bouchand},\ and\ \citenamefont {Kippenberg}}]{Liu:19}%
  \BibitemOpen
  \bibfield  {author} {\bibinfo {author} {\bibfnamefont {J.}~\bibnamefont
  {Liu}}, \bibinfo {author} {\bibfnamefont {E.}~\bibnamefont {Lucas}}, \bibinfo
  {author} {\bibfnamefont {J.}~\bibnamefont {He}}, \bibinfo {author}
  {\bibfnamefont {A.~S.}\ \bibnamefont {Raja}}, \bibinfo {author}
  {\bibfnamefont {R.~N.}\ \bibnamefont {Wang}}, \bibinfo {author}
  {\bibfnamefont {M.}~\bibnamefont {Karpov}}, \bibinfo {author} {\bibfnamefont
  {H.}~\bibnamefont {Guo}}, \bibinfo {author} {\bibfnamefont {R.}~\bibnamefont
  {Bouchand}}, \ and\ \bibinfo {author} {\bibfnamefont {T.~J.}\ \bibnamefont
  {Kippenberg}},\ }\href {http://arxiv.org/abs/1901.10372} {\bibfield
  {journal} {\bibinfo  {journal} {arXiv:1901.10372 [physics]}\ } (\bibinfo
  {year} {2019})}\BibitemShut {NoStop}%
\bibitem [{\citenamefont {Brasch}\ \emph {et~al.}(2016)\citenamefont {Brasch},
  \citenamefont {Geiselmann}, \citenamefont {Pfeiffer},\ and\ \citenamefont
  {Kippenberg}}]{Brasch:16}%
  \BibitemOpen
  \bibfield  {author} {\bibinfo {author} {\bibfnamefont {V.}~\bibnamefont
  {Brasch}}, \bibinfo {author} {\bibfnamefont {M.}~\bibnamefont {Geiselmann}},
  \bibinfo {author} {\bibfnamefont {M.~H.~P.}\ \bibnamefont {Pfeiffer}}, \ and\
  \bibinfo {author} {\bibfnamefont {T.~J.}\ \bibnamefont {Kippenberg}},\ }\href
  {\doibase 10.1364/OE.24.029312} {\bibfield  {journal} {\bibinfo  {journal}
  {Opt. Express}\ }\textbf {\bibinfo {volume} {24}},\ \bibinfo {pages} {29312}
  (\bibinfo {year} {2016})}\BibitemShut {NoStop}%
\bibitem [{\citenamefont {Stone}\ \emph {et~al.}(2018)\citenamefont {Stone},
  \citenamefont {Briles}, \citenamefont {Drake}, \citenamefont {Spencer},
  \citenamefont {Carlson}, \citenamefont {Diddams},\ and\ \citenamefont
  {Papp}}]{Stone:18}%
  \BibitemOpen
  \bibfield  {author} {\bibinfo {author} {\bibfnamefont {J.~R.}\ \bibnamefont
  {Stone}}, \bibinfo {author} {\bibfnamefont {T.~C.}\ \bibnamefont {Briles}},
  \bibinfo {author} {\bibfnamefont {T.~E.}\ \bibnamefont {Drake}}, \bibinfo
  {author} {\bibfnamefont {D.~T.}\ \bibnamefont {Spencer}}, \bibinfo {author}
  {\bibfnamefont {D.~R.}\ \bibnamefont {Carlson}}, \bibinfo {author}
  {\bibfnamefont {S.~A.}\ \bibnamefont {Diddams}}, \ and\ \bibinfo {author}
  {\bibfnamefont {S.~B.}\ \bibnamefont {Papp}},\ }\href {\doibase
  10.1103/PhysRevLett.121.063902} {\bibfield  {journal} {\bibinfo  {journal}
  {Phys. Rev. Lett.}\ }\textbf {\bibinfo {volume} {121}},\ \bibinfo {pages}
  {063902} (\bibinfo {year} {2018})}\BibitemShut {NoStop}%
\bibitem [{\citenamefont {Yang}\ \emph {et~al.}(2018)\citenamefont {Yang},
  \citenamefont {Oh}, \citenamefont {Lee}, \citenamefont {Yang}, \citenamefont
  {Yi}, \citenamefont {Shen}, \citenamefont {Wang},\ and\ \citenamefont
  {Vahala}}]{Yang:18}%
  \BibitemOpen
  \bibfield  {author} {\bibinfo {author} {\bibfnamefont {K.~Y.}\ \bibnamefont
  {Yang}}, \bibinfo {author} {\bibfnamefont {D.~Y.}\ \bibnamefont {Oh}},
  \bibinfo {author} {\bibfnamefont {S.~H.}\ \bibnamefont {Lee}}, \bibinfo
  {author} {\bibfnamefont {Q.-F.}\ \bibnamefont {Yang}}, \bibinfo {author}
  {\bibfnamefont {X.}~\bibnamefont {Yi}}, \bibinfo {author} {\bibfnamefont
  {B.}~\bibnamefont {Shen}}, \bibinfo {author} {\bibfnamefont {H.}~\bibnamefont
  {Wang}}, \ and\ \bibinfo {author} {\bibfnamefont {K.}~\bibnamefont
  {Vahala}},\ }\href {\doibase 10.1038/s41566-018-0132-5} {\bibfield  {journal}
  {\bibinfo  {journal} {Nature Photonics}\ }\textbf {\bibinfo {volume} {12}},\
  \bibinfo {pages} {297} (\bibinfo {year} {2018})}\BibitemShut {NoStop}%
\bibitem [{\citenamefont {Stern}\ \emph {et~al.}(2018)\citenamefont {Stern},
  \citenamefont {Ji}, \citenamefont {Okawachi}, \citenamefont {Gaeta},\ and\
  \citenamefont {Lipson}}]{Stern:18}%
  \BibitemOpen
  \bibfield  {author} {\bibinfo {author} {\bibfnamefont {B.}~\bibnamefont
  {Stern}}, \bibinfo {author} {\bibfnamefont {X.}~\bibnamefont {Ji}}, \bibinfo
  {author} {\bibfnamefont {Y.}~\bibnamefont {Okawachi}}, \bibinfo {author}
  {\bibfnamefont {A.~L.}\ \bibnamefont {Gaeta}}, \ and\ \bibinfo {author}
  {\bibfnamefont {M.}~\bibnamefont {Lipson}},\ }\href {\doibase
  10.1038/s41586-018-0598-9} {\bibfield  {journal} {\bibinfo  {journal}
  {Nature}\ }\textbf {\bibinfo {volume} {562}},\ \bibinfo {pages} {401}
  (\bibinfo {year} {2018})}\BibitemShut {NoStop}%
\bibitem [{\citenamefont {Raja}\ \emph {et~al.}(2019)\citenamefont {Raja},
  \citenamefont {Voloshin}, \citenamefont {Guo}, \citenamefont {Agafonova},
  \citenamefont {Liu}, \citenamefont {Gorodnitskiy}, \citenamefont {Karpov},
  \citenamefont {Pavlov}, \citenamefont {Lucas}, \citenamefont {Galiev},
  \citenamefont {Shitikov}, \citenamefont {Jost}, \citenamefont {Gorodetsky},\
  and\ \citenamefont {Kippenberg}}]{Raja:19}%
  \BibitemOpen
  \bibfield  {author} {\bibinfo {author} {\bibfnamefont {A.~S.}\ \bibnamefont
  {Raja}}, \bibinfo {author} {\bibfnamefont {A.~S.}\ \bibnamefont {Voloshin}},
  \bibinfo {author} {\bibfnamefont {H.}~\bibnamefont {Guo}}, \bibinfo {author}
  {\bibfnamefont {S.~E.}\ \bibnamefont {Agafonova}}, \bibinfo {author}
  {\bibfnamefont {J.}~\bibnamefont {Liu}}, \bibinfo {author} {\bibfnamefont
  {A.~S.}\ \bibnamefont {Gorodnitskiy}}, \bibinfo {author} {\bibfnamefont
  {M.}~\bibnamefont {Karpov}}, \bibinfo {author} {\bibfnamefont {N.~G.}\
  \bibnamefont {Pavlov}}, \bibinfo {author} {\bibfnamefont {E.}~\bibnamefont
  {Lucas}}, \bibinfo {author} {\bibfnamefont {R.~R.}\ \bibnamefont {Galiev}},
  \bibinfo {author} {\bibfnamefont {A.~E.}\ \bibnamefont {Shitikov}}, \bibinfo
  {author} {\bibfnamefont {J.~D.}\ \bibnamefont {Jost}}, \bibinfo {author}
  {\bibfnamefont {M.~L.}\ \bibnamefont {Gorodetsky}}, \ and\ \bibinfo {author}
  {\bibfnamefont {T.~J.}\ \bibnamefont {Kippenberg}},\ }\href {\doibase
  10.1038/s41467-019-08498-2} {\bibfield  {journal} {\bibinfo  {journal}
  {Nature Communications}\ }\textbf {\bibinfo {volume} {10}},\ \bibinfo {pages}
  {680} (\bibinfo {year} {2019})}\BibitemShut {NoStop}%
\bibitem [{\citenamefont {Morton}\ and\ \citenamefont
  {Morton}(2018)}]{Morton:18}%
  \BibitemOpen
  \bibfield  {author} {\bibinfo {author} {\bibfnamefont {P.~A.}\ \bibnamefont
  {Morton}}\ and\ \bibinfo {author} {\bibfnamefont {M.~J.}\ \bibnamefont
  {Morton}},\ }\href {http://jlt.osa.org/abstract.cfm?URI=jlt-36-21-5048}
  {\bibfield  {journal} {\bibinfo  {journal} {J. Lightwave Technol.}\ }\textbf
  {\bibinfo {volume} {36}},\ \bibinfo {pages} {5048} (\bibinfo {year}
  {2018})}\BibitemShut {NoStop}%
\bibitem [{\citenamefont {Pfeiffer}\ \emph
  {et~al.}(2018{\natexlab{a}})\citenamefont {Pfeiffer}, \citenamefont {Liu},
  \citenamefont {Raja}, \citenamefont {Morais}, \citenamefont {Ghadiani},\ and\
  \citenamefont {Kippenberg}}]{Pfeiffer:18}%
  \BibitemOpen
  \bibfield  {author} {\bibinfo {author} {\bibfnamefont {M.~H.~P.}\
  \bibnamefont {Pfeiffer}}, \bibinfo {author} {\bibfnamefont {J.}~\bibnamefont
  {Liu}}, \bibinfo {author} {\bibfnamefont {A.~S.}\ \bibnamefont {Raja}},
  \bibinfo {author} {\bibfnamefont {T.}~\bibnamefont {Morais}}, \bibinfo
  {author} {\bibfnamefont {B.}~\bibnamefont {Ghadiani}}, \ and\ \bibinfo
  {author} {\bibfnamefont {T.~J.}\ \bibnamefont {Kippenberg}},\ }\href
  {\doibase 10.1364/OPTICA.5.000884} {\bibfield  {journal} {\bibinfo  {journal}
  {Optica}\ }\textbf {\bibinfo {volume} {5}},\ \bibinfo {pages} {884} (\bibinfo
  {year} {2018}{\natexlab{a}})}\BibitemShut {NoStop}%
\bibitem [{\citenamefont {Liu}\ \emph {et~al.}(2018{\natexlab{a}})\citenamefont
  {Liu}, \citenamefont {Raja}, \citenamefont {Karpov}, \citenamefont
  {Ghadiani}, \citenamefont {Pfeiffer}, \citenamefont {Du}, \citenamefont
  {Engelsen}, \citenamefont {Guo}, \citenamefont {Zervas},\ and\ \citenamefont
  {Kippenberg}}]{Liu:18a}%
  \BibitemOpen
  \bibfield  {author} {\bibinfo {author} {\bibfnamefont {J.}~\bibnamefont
  {Liu}}, \bibinfo {author} {\bibfnamefont {A.~S.}\ \bibnamefont {Raja}},
  \bibinfo {author} {\bibfnamefont {M.}~\bibnamefont {Karpov}}, \bibinfo
  {author} {\bibfnamefont {B.}~\bibnamefont {Ghadiani}}, \bibinfo {author}
  {\bibfnamefont {M.~H.~P.}\ \bibnamefont {Pfeiffer}}, \bibinfo {author}
  {\bibfnamefont {B.}~\bibnamefont {Du}}, \bibinfo {author} {\bibfnamefont
  {N.~J.}\ \bibnamefont {Engelsen}}, \bibinfo {author} {\bibfnamefont
  {H.}~\bibnamefont {Guo}}, \bibinfo {author} {\bibfnamefont {M.}~\bibnamefont
  {Zervas}}, \ and\ \bibinfo {author} {\bibfnamefont {T.~J.}\ \bibnamefont
  {Kippenberg}},\ }\href {\doibase 10.1364/OPTICA.5.001347} {\bibfield
  {journal} {\bibinfo  {journal} {Optica}\ }\textbf {\bibinfo {volume} {5}},\
  \bibinfo {pages} {1347} (\bibinfo {year} {2018}{\natexlab{a}})}\BibitemShut
  {NoStop}%
\bibitem [{\citenamefont {Volet}\ \emph {et~al.}(2018)\citenamefont {Volet},
  \citenamefont {Yi}, \citenamefont {Yang}, \citenamefont {Stanton},
  \citenamefont {Morton}, \citenamefont {Yang}, \citenamefont {Vahala},\ and\
  \citenamefont {Bowers}}]{Nick:18}%
  \BibitemOpen
  \bibfield  {author} {\bibinfo {author} {\bibfnamefont {N.}~\bibnamefont
  {Volet}}, \bibinfo {author} {\bibfnamefont {X.}~\bibnamefont {Yi}}, \bibinfo
  {author} {\bibfnamefont {Q.-F.}\ \bibnamefont {Yang}}, \bibinfo {author}
  {\bibfnamefont {E.~J.}\ \bibnamefont {Stanton}}, \bibinfo {author}
  {\bibfnamefont {P.~A.}\ \bibnamefont {Morton}}, \bibinfo {author}
  {\bibfnamefont {K.~Y.}\ \bibnamefont {Yang}}, \bibinfo {author}
  {\bibfnamefont {K.~J.}\ \bibnamefont {Vahala}}, \ and\ \bibinfo {author}
  {\bibfnamefont {J.~E.}\ \bibnamefont {Bowers}},\ }\href {\doibase
  10.1002/lpor.201700307} {\bibfield  {journal} {\bibinfo  {journal} {Laser \&
  Photonics Reviews}\ }\textbf {\bibinfo {volume} {12}},\ \bibinfo {pages}
  {1700307} (\bibinfo {year} {2018})}\BibitemShut {NoStop}%
\bibitem [{\citenamefont {Suh}\ \emph {et~al.}(2019{\natexlab{b}})\citenamefont
  {Suh}, \citenamefont {Wang}, \citenamefont {Johnson},\ and\ \citenamefont
  {Vahala}}]{Suh:19}%
  \BibitemOpen
  \bibfield  {author} {\bibinfo {author} {\bibfnamefont {M.-G.}\ \bibnamefont
  {Suh}}, \bibinfo {author} {\bibfnamefont {C.~Y.}\ \bibnamefont {Wang}},
  \bibinfo {author} {\bibfnamefont {C.}~\bibnamefont {Johnson}}, \ and\
  \bibinfo {author} {\bibfnamefont {K.}~\bibnamefont {Vahala}},\ }\href
  {http://arxiv.org/abs/1901.08126} {\bibfield  {journal} {\bibinfo  {journal}
  {arXiv:1901.08126 [physics]}\ } (\bibinfo {year}
  {2019}{\natexlab{b}})}\BibitemShut {NoStop}%
\bibitem [{\citenamefont {Pfeiffer}\ \emph
  {et~al.}(2018{\natexlab{b}})\citenamefont {Pfeiffer}, \citenamefont
  {Herkommer}, \citenamefont {Liu}, \citenamefont {Morais}, \citenamefont
  {Zervas}, \citenamefont {Geiselmann},\ and\ \citenamefont
  {Kippenberg}}]{Pfeiffer:18a}%
  \BibitemOpen
  \bibfield  {author} {\bibinfo {author} {\bibfnamefont {M.~H.~P.}\
  \bibnamefont {Pfeiffer}}, \bibinfo {author} {\bibfnamefont {C.}~\bibnamefont
  {Herkommer}}, \bibinfo {author} {\bibfnamefont {J.}~\bibnamefont {Liu}},
  \bibinfo {author} {\bibfnamefont {T.}~\bibnamefont {Morais}}, \bibinfo
  {author} {\bibfnamefont {M.}~\bibnamefont {Zervas}}, \bibinfo {author}
  {\bibfnamefont {M.}~\bibnamefont {Geiselmann}}, \ and\ \bibinfo {author}
  {\bibfnamefont {T.~J.}\ \bibnamefont {Kippenberg}},\ }\bibfield  {booktitle}
  {\emph {\bibinfo {booktitle} {IEEE Journal of Selected Topics in Quantum
  Electronics}},\ }\href {\doibase 10.1109/JSTQE.2018.2808258} {\bibfield
  {journal} {\bibinfo  {journal} {IEEE Journal of Selected Topics in Quantum
  Electronics}\ }\textbf {\bibinfo {volume} {24}},\ \bibinfo {pages} {1}
  (\bibinfo {year} {2018}{\natexlab{b}})}\BibitemShut {NoStop}%
\bibitem [{\citenamefont {Pfeiffer}\ \emph {et~al.}(2017)\citenamefont
  {Pfeiffer}, \citenamefont {Liu}, \citenamefont {Geiselmann},\ and\
  \citenamefont {Kippenberg}}]{Pfeiffer:17b}%
  \BibitemOpen
  \bibfield  {author} {\bibinfo {author} {\bibfnamefont {M.~H.~P.}\
  \bibnamefont {Pfeiffer}}, \bibinfo {author} {\bibfnamefont {J.}~\bibnamefont
  {Liu}}, \bibinfo {author} {\bibfnamefont {M.}~\bibnamefont {Geiselmann}}, \
  and\ \bibinfo {author} {\bibfnamefont {T.~J.}\ \bibnamefont {Kippenberg}},\
  }\href {\doibase 10.1103/PhysRevApplied.7.024026} {\bibfield  {journal}
  {\bibinfo  {journal} {Phys. Rev. Applied}\ }\textbf {\bibinfo {volume} {7}},\
  \bibinfo {pages} {024026} (\bibinfo {year} {2017})}\BibitemShut {NoStop}%
\bibitem [{\citenamefont {Liu}\ \emph {et~al.}(2018{\natexlab{b}})\citenamefont
  {Liu}, \citenamefont {Raja}, \citenamefont {Pfeiffer}, \citenamefont
  {Herkommer}, \citenamefont {Guo}, \citenamefont {Zervas}, \citenamefont
  {Geiselmann},\ and\ \citenamefont {Kippenberg}}]{Liu:18}%
  \BibitemOpen
  \bibfield  {author} {\bibinfo {author} {\bibfnamefont {J.}~\bibnamefont
  {Liu}}, \bibinfo {author} {\bibfnamefont {A.~S.}\ \bibnamefont {Raja}},
  \bibinfo {author} {\bibfnamefont {M.~H.~P.}\ \bibnamefont {Pfeiffer}},
  \bibinfo {author} {\bibfnamefont {C.}~\bibnamefont {Herkommer}}, \bibinfo
  {author} {\bibfnamefont {H.}~\bibnamefont {Guo}}, \bibinfo {author}
  {\bibfnamefont {M.}~\bibnamefont {Zervas}}, \bibinfo {author} {\bibfnamefont
  {M.}~\bibnamefont {Geiselmann}}, \ and\ \bibinfo {author} {\bibfnamefont
  {T.~J.}\ \bibnamefont {Kippenberg}},\ }\href {\doibase 10.1364/OL.43.003200}
  {\bibfield  {journal} {\bibinfo  {journal} {Opt. Lett.}\ }\textbf {\bibinfo
  {volume} {43}},\ \bibinfo {pages} {3200} (\bibinfo {year}
  {2018}{\natexlab{b}})}\BibitemShut {NoStop}%
\bibitem [{\citenamefont {Del'Haye}\ \emph {et~al.}(2009)\citenamefont
  {Del'Haye}, \citenamefont {Arcizet}, \citenamefont {Gorodetsky},
  \citenamefont {Holzwarth},\ and\ \citenamefont {Kippenberg}}]{DelHaye:09}%
  \BibitemOpen
  \bibfield  {author} {\bibinfo {author} {\bibfnamefont {P.}~\bibnamefont
  {Del'Haye}}, \bibinfo {author} {\bibfnamefont {O.}~\bibnamefont {Arcizet}},
  \bibinfo {author} {\bibfnamefont {M.~L.}\ \bibnamefont {Gorodetsky}},
  \bibinfo {author} {\bibfnamefont {R.}~\bibnamefont {Holzwarth}}, \ and\
  \bibinfo {author} {\bibfnamefont {T.~J.}\ \bibnamefont {Kippenberg}},\ }\href
  {http://dx.doi.org/10.1038/nphoton.2009.138} {\bibfield  {journal} {\bibinfo
  {journal} {Nature Photonics}\ }\textbf {\bibinfo {volume} {3}},\ \bibinfo
  {pages} {529} (\bibinfo {year} {2009})}\BibitemShut {NoStop}%
\bibitem [{\citenamefont {Liu}\ \emph {et~al.}(2016)\citenamefont {Liu},
  \citenamefont {Brasch}, \citenamefont {Pfeiffer}, \citenamefont {Kordts},
  \citenamefont {Kamel}, \citenamefont {Guo}, \citenamefont {Geiselmann},\ and\
  \citenamefont {Kippenberg}}]{Liu:16}%
  \BibitemOpen
  \bibfield  {author} {\bibinfo {author} {\bibfnamefont {J.}~\bibnamefont
  {Liu}}, \bibinfo {author} {\bibfnamefont {V.}~\bibnamefont {Brasch}},
  \bibinfo {author} {\bibfnamefont {M.~H.~P.}\ \bibnamefont {Pfeiffer}},
  \bibinfo {author} {\bibfnamefont {A.}~\bibnamefont {Kordts}}, \bibinfo
  {author} {\bibfnamefont {A.~N.}\ \bibnamefont {Kamel}}, \bibinfo {author}
  {\bibfnamefont {H.}~\bibnamefont {Guo}}, \bibinfo {author} {\bibfnamefont
  {M.}~\bibnamefont {Geiselmann}}, \ and\ \bibinfo {author} {\bibfnamefont
  {T.~J.}\ \bibnamefont {Kippenberg}},\ }\href {\doibase 10.1364/OL.41.003134}
  {\bibfield  {journal} {\bibinfo  {journal} {Opt. Lett.}\ }\textbf {\bibinfo
  {volume} {41}},\ \bibinfo {pages} {3134} (\bibinfo {year}
  {2016})}\BibitemShut {NoStop}%
\bibitem [{\citenamefont {Gorodetsky}\ \emph {et~al.}(2000)\citenamefont
  {Gorodetsky}, \citenamefont {Pryamikov},\ and\ \citenamefont
  {Ilchenko}}]{Gorodetsky:00}%
  \BibitemOpen
  \bibfield  {author} {\bibinfo {author} {\bibfnamefont {M.~L.}\ \bibnamefont
  {Gorodetsky}}, \bibinfo {author} {\bibfnamefont {A.~D.}\ \bibnamefont
  {Pryamikov}}, \ and\ \bibinfo {author} {\bibfnamefont {V.~S.}\ \bibnamefont
  {Ilchenko}},\ }\href {\doibase 10.1364/JOSAB.17.001051} {\bibfield  {journal}
  {\bibinfo  {journal} {J. Opt. Soc. Am. B}\ }\textbf {\bibinfo {volume}
  {17}},\ \bibinfo {pages} {1051} (\bibinfo {year} {2000})}\BibitemShut
  {NoStop}%
\bibitem [{\citenamefont {Zhang}\ \emph {et~al.}(2019)\citenamefont {Zhang},
  \citenamefont {Silver}, \citenamefont {Bino}, \citenamefont {Copie},
  \citenamefont {Woodley}, \citenamefont {Ghalanos}, \citenamefont {Svela},
  \citenamefont {Moroney},\ and\ \citenamefont {Del'Haye}}]{Zhang:19}%
  \BibitemOpen
  \bibfield  {author} {\bibinfo {author} {\bibfnamefont {S.}~\bibnamefont
  {Zhang}}, \bibinfo {author} {\bibfnamefont {J.~M.}\ \bibnamefont {Silver}},
  \bibinfo {author} {\bibfnamefont {L.~D.}\ \bibnamefont {Bino}}, \bibinfo
  {author} {\bibfnamefont {F.}~\bibnamefont {Copie}}, \bibinfo {author}
  {\bibfnamefont {M.~T.~M.}\ \bibnamefont {Woodley}}, \bibinfo {author}
  {\bibfnamefont {G.~N.}\ \bibnamefont {Ghalanos}}, \bibinfo {author}
  {\bibfnamefont {A.~{\O}.}\ \bibnamefont {Svela}}, \bibinfo {author}
  {\bibfnamefont {N.}~\bibnamefont {Moroney}}, \ and\ \bibinfo {author}
  {\bibfnamefont {P.}~\bibnamefont {Del'Haye}},\ }\href {\doibase
  10.1364/OPTICA.6.000206} {\bibfield  {journal} {\bibinfo  {journal} {Optica}\
  }\textbf {\bibinfo {volume} {6}},\ \bibinfo {pages} {206} (\bibinfo {year}
  {2019})}\BibitemShut {NoStop}%
\end{thebibliography}%
\end{document}